\documentclass[12pt]{JHEP3}

\usepackage[pdftex]{graphicx}
\newcommand\fverb{\setbox\pippobox=\hbox\bgroup\verb}
\newcommand\fverbdo{\egroup\medskip\noindent%
\fbox{\unhbox\pippobox}\ }                      
\newcommand\fverbit{\egroup\item[\fbox{\unhbox\pippobox}]}
\newbox\pippobox
\def\d2bar{$\overline{\mbox D2}$}
\title{The Entropy Function for the extremal Kerr-(anti-)de Sitter Black Holes}
\author{Jin-Ho Cho$^{1,2}$, Yumi Ko$^{1}$ and Soonkeon Nam$^{1}$\\
$^{1}$Department of Physics \& Research Institute for Basic Sciences,\\ Kyung Hee University, Seoul 130-701, Korea\\
$^{2}$Center for Quantum Space Time, Sogang University, Seoul 121-742, Korea\\\\
E-mail: \email{cho.jinho@gmail.com}, \email{koyumi@khu.ac.kr}, \email{nam@khu.ac.kr}}

\preprint{arXiv{:0804.3811}} 

\abstract{Based on the entropy function formalism, we consider the extremal Kerr-(anti-)de Sitter black holes in $4$-dimensions. Solving differential equations exactly, which are obtained by extremizing the entropy function, we find agreement of the result with Bekenstein-Hawking entropy. Concerning the higher derivative corrections, we extend the computation to the case with Gauss-Bonnet term. 
}  


\begin{document}

\section{Introduction}

The Bekenstein-Hawking entropy of black holes is a gateway to the quantum theory of gravity \cite{{Bekenstein:1973ur},{Hawking:1974sw}}. Although the entropy is given by a simple expression, $S_{BH} = A/4G_N$, where $A$ is the area of the black hole event horizon, we do not have a complete understanding of its microscopic origin. 

String theory gives a good microscopic description but its low energy effective theory gives rise to many additional fields as well as higher order terms for the gravity part \cite{Callan:1985ia}. The entropy of a general black hole in the context of the higher order gravity is hard to compute even macroscopically when we do not have  a concrete form of the solution. 

Recently, this problem has been remedied by the entropy function formalism of Ashoke Sen for the extremal black holes \cite{{Sen:2005wa},{Sen:2005iz}}. This allows us to compute the Bekenstein-Hawking entropy without full knowledge about the metric. The only necessary information is the near horizon isometry of a black hole. The extremal Reissner-Nordstr\"{o}m black hole in 4-dimensions, which has the isometry $SO(1,2)\times SO(3)$, is a simple example where the entropy function formalism is applicable.  
There have been several works on non-supersymmetric extremal black holes. 
The examples include charged black holes in de Sitter background \cite{Cho:2007mn}, rotating black holes \cite{{Astefanesei:2006dd},{Astefanesei:2007bf},{Cardoso:2007rg}}, and black holes of Nariai class \cite{Cho:2007we}. In these cases, the entropy function formalism gives the correct values of the Bekenstein-Hawking entropy.

In this paper, we would like to study yet another example, that of rotating black holes {\it{in asymptotically (anti-)de Sitter space-time}} in 4-dimensions. We are generalizing the case of rotating black holes in the asymptotically flat space-time \cite{Astefanesei:2006dd}. We will see that the entropy function formalism is also applicable in the Kerr-(anti-)de Sitter black hole case. We will also see the effect of Gauss-Bonnet term in the Lagrangian on the black hole entropy. 

Entropy function formalism is not easy to apply for the rotating black holes. It is because the near horizon geometry of the rotating black hole is no longer two dimensional anti-de Sitter space-time\,(\,AdS$_2$) times a round two-sphere\,(\,S$^2$). Rather, the near horizon geometry has $SO(1,2)\times U(1)$ isometry, and both the AdS$_2$ part and the circle\,(\,S$^1$) part are warped by a function depending on the latitudinal angle $\theta$.  A rotating black hole has a horizon whose shape is a smooth deformation of a round sphere. Unlike static cases, metric coefficients are now functions of $\theta$. Despite this difficulty, the entropy function formalism is still valid for the rotating black holes in the flat background, as long as the near horizon geometry contains AdS$_2$ part \cite{Astefanesei:2006dd}.  

It is meaningful to check whether the entropy function formalism works for the rotating black holes in (anti-)de Sitter background. These black holes are more realistic in the sense that we are now in an accelerating universe and most of the existing black holes are rotating. Exploring Kerr-de Sitter case is also important on the theoretical side. It provides another example of non-supersymmetric attractor\footnote{It is an attractor in the sense that the physical quantities like the entropy are independent of the asymptotic boundary structure of the geometry.}. Regarding the non-supersymmetric attractor, there have been many studies on non-BPS but extremal black holes in the supersymmetric background \cite{{Goldstein:2005hq},{Tripathy:2005qp},{Nampuri:2007gv}}. Reissner-Nordstr\"{o}m-de Sitter black hole was studied in  Ref.\cite{Cho:2007mn}, as the first example showing the feature of non-supersymmetric attractor in non-supersymmetric vacuum. 

We organize our paper as follows: Sec. 2 discusses general properties of the extremal Kerr-(anti-)de Sitter black hole. A relation between Bekenstein-Hawking entropy $S$, and the angular momentum $J$, is obtained (See Eq.(\ref{r1})). The near horizon geometry of the extremal black hole in (anti-)de Sitter spacetime contains  AdS$_2$ symmetry. In Sec. 3, making an ansatz for the near horizon geometry of the extremal Kerr-(anti-)de Sitter black hole, we write down the entropy function. We solve the `Euler-Lagrange equations' minimizing the entropy function  and obtain the black hole entropy as an extremum value of the entropy function. We verify the entropy function formalism for the case at hand, by recovering the relation of Eq.(\ref{r1}) between the entropy and the angular momentum. (See Eq.(\ref{algerel})). We then apply this method to obtain higher order corrections to the entropy. The case with Gauss-Bonnet term is considered.

\section{Kerr-(anti-)de Sitter Black Holes } 
In this section, we consider the general properties of extremal Kerr-(anti-)de Sitter black holes such as the horizon structure, Hawking temperature, and the entropy. In particular, we obtain a relation between the Bekenstein-Hawking entropy and the angular momentum. The near horizon geometry of these black holes has $SO(1,2) \times U(1)$ isometry. Using this isometry, we will write down an ansatz for the near horizon geometry of rotating black holes in asymptotically (anti-)de Sitter spacetime in the next section.

\subsection{Preliminary : Four Dimensional Kerr Black Holes}

Prior to our main issues on the Kerr-(anti-)de Sitter black holes, we will warm up our discussion by considering the basic  properties of the extremal Kerr black hole.
We will observe that the rotation gives the warping of AdS$_2$ and $S^1$ by functions depending on $\theta$. 

Four Dimensional Kerr metric in Boyer-Lindquist form is given as \cite{Boyer:1966qh}
\begin{equation}\label{kerr}
ds^2 = -\frac{\Delta}{\rho^2}( dt - a \sin^2 \theta d\phi)^2 +\frac{\rho^2}{\Delta} dr^2 + \rho^2 d \theta^2 + \frac{\sin^2 \theta}{\rho^2} \biggl( (r^2+a^2) d\phi - a dt \biggr)^2,
\end{equation}
where
\begin{eqnarray}
\rho^2 = r^2 + a^2 \cos^2 \theta, \quad \mbox{and}\quad \Delta = r^2 - 2 m r + a^2.
\end{eqnarray}
Here, the angular momentum parameter $a=J/m$, where $m$ denotes ADM mass and $J$ is angular momentum. We use the geometrical unit that sets $c=1$ and $G_N = 1$. With non vanishing angular momentum $( a\neq 0)$ the metric function is not spherically symmetric.

The near horizon geometry of the extremal case contains the universal factor AdS$_2$.  
When $ r_0^2=a^2 = m^2 $, we obtain the extremal Kerr black hole with the event horizon at $r=r_0$. Now let us introduce the near horizon coordinates $(\bar{t}, \bar{r}, \bar{\phi})$  as 
\begin{equation}
{ r}=r_0 + \epsilon {\bar r}, ~~~ { t} = \frac{\bar t}{ \epsilon},~~~{ \phi} = {\bar\phi}+\frac{\bar t}{2 r_0 \epsilon},
 \end{equation}
where $\epsilon$ is a dimensionless parameter. By shifting $\phi$ to ${\bar \phi}$, \,$\partial/ \partial {\bar t}$ ~becomes null at the horizon ${\bar r}=0$. It implies that the coordinate co-rotates with the horizon \cite{Bardeen:1999px}. 
By taking a limit $\epsilon\rightarrow 0$, we can obtain the near horizon geometry of the extremal Kerr black holes as
\begin{eqnarray}\label{nhk}
ds^2 = \biggl( \frac{1+ \cos^2 \theta}{2} \biggr) \biggl( -\frac{ {\bar r}^2 }{ 2 r_0^2} d{\bar t}^2 +\frac { 2 r_0^2}{ {\bar r}^2} d{\bar r}^2 +2 r_0^2 d \theta^2\biggr) +\frac{4 r_0^2 \sin^2 \theta}{1+ \cos^2 \theta}\biggl( d {\bar \phi} + \frac{\bar r}{2 r_0^2} d{\bar t}\biggr)^{\!2}\!. ~~~~
\end{eqnarray}
By rescaling ${\bar r}$ as
\begin{equation}
{\tilde r} = \frac{\bar r}{ 2 r_0^2},
\end{equation}
we can rewrite Eq.(\ref{nhk}) in a form that shows manifestly ${\rm AdS}_2 \times {\rm S}^1$ structures but warped by functions depending on $\theta$;  
\begin{equation}\label{nhkerr}
ds^2 = r_0^2 (1 + \cos^2 \theta)\biggl(- {\tilde r}^2 d{\bar t}^2 + \frac{d{\tilde r}^2}{{\tilde r}^2}  + d \theta^2 \biggr) + \frac{4 r_0^2 \sin^2 \theta}{1+ \cos^2 \theta} \left( d {\bar\phi} + {\tilde r} d{\bar t}\,\right)^2.
\end{equation}

\subsection{Horizon Structure of Kerr-(anti-)de Sitter Black Holes}
In $d$-dimensional gravity theories with cosmological constant, we have Kerr-(anti-) de Sitter black holes, that is, a generic form of the stationary black hole solutions. There are two kinds of parameters characterizing the solutions; the parameter $a_{i}$ ($i=1,2,\cdots$) concerning various rotations, and the length scale $l$  parametrizing the size of the (anti-)de Sitter spacetime as
\begin{eqnarray}
\frac{\eta}{l^2} = \frac{2 }{(d-1)(d-2)} \Lambda_d, \qquad\quad (d\ge 3).
\end{eqnarray}
Here, we have $\eta = +1/ -1$ for de Sitter/anti-de Sitter respectively\footnote{Introducing $\eta$, we can have unified expressions for both de Sitter and anti-de Sitter cases. There is a big difference, however, when we consider the horizon structure. For example, only for de Sitter case, we have a cosmological horizon.}. 

Kerr-(anti-)de Sitter solutions in various dimensions are summarized as follows \cite{{Carter:1968ks},{Dehghani:2001kn}}: 
\begin{eqnarray}\label{kds1}
ds^2 &=& -\frac{\Delta_r}{\rho^2}\biggl(d{ t} - \frac{a}{\Xi} \sin^2 \theta d{\phi} \biggr)^2 + \frac{\rho^2}{\Delta_r}\, d{r}^2 + \frac{\rho^2}{\Delta_\theta}\, d\theta^2 \nonumber\\
&&+\frac{{\Delta_\theta}\sin^2 \theta}{\rho^2} \biggl( a\, d{ t} - \frac{({r}^2 + a^2 )}{\Xi}
d{\phi} \biggr)^2
 + r^2 \cos^2 \theta \,d\Omega_{d-4},
\end{eqnarray}
where
\begin{eqnarray}\label{kds2}
\Delta_r &\equiv& \Delta_r (r)= ( {r}^2 + a^2 )\left(1-\frac{ \eta}{l^2 }r^2\right) - 2 m {r^{5-d}},\nonumber\\
\Delta_\theta&=& 1 + \frac{\eta}{ l^2 } a^2 \cos^2 \theta,\nonumber\\
\Xi&=& 1 + \frac{\eta}{ l^2 }a^2,\nonumber\\
\rho^2 &=& {r}^2 + a^2 \cos^2 \theta.
\end{eqnarray}
We note that the solution approaches Kerr solution as $l \rightarrow \infty$~ or Schwarzschild-(anti-)de Sitter solution when ~$a \rightarrow 0$. 
Here, for simplicity, we just consider the case with a single parameter $a$ for the rotation. It is assumed to satisfy $a^2 <l^2$ so that $\Delta_{\theta} >0$ for any value of $\theta$. When the cosmological constant is non-vanishing, one cannot give physical meaning to the parameters $a$ and $m$ directly. The meaning will be given below. 

In the case of Kerr-(anti-) de Sitter black holes in 4-dimensions, the conserved charges (in the spirit of Abbott and Deser \cite{Abbott:1981ff}) of a rotating black hole are  the `mass', $M$, and the `angular momentum', $J$. These quantities, associated with the black hole event horizon, are related with the parameters $m$, $a$ and $l$ (of Eqs.(\ref{kds1}) and (\ref{kds2})) as \cite{{Cai:2001tv},{Hawking:1998kw}};
\begin{eqnarray}\label{mj}
M= \frac{m}{ \Xi}, \quad J= \frac{m a}{\Xi^2}.
\end{eqnarray}
One can check that the `mass' $M$ coincides with the ADM mass in the asymptotically flat spacetime as by limiting $l \rightarrow \infty$.

In the rest of this paper, we will focus on the 4-dimensional case only, for simplicity.
The horizons are located at the zeros of $\Delta_r(r)$. We will draw the form of the function $\Delta_{r}(r)$. A simple analysis on the derivative of the function $\Delta_{r}(r)$ shows that it has at most three extremal values in de Sitter case ($\eta=1$) as in Fig. \ref{Gmass1}, while it allows only one in anti-de Sitter case ($\eta=-1$) as in Fig. \ref{Gmass2}. 

In de Sitter case, the generic form of the function $\Delta_r(r)$ can be written as follows:
\begin{eqnarray}
\Delta_r (r)=  -\frac{1}{l^2}( r- r_-)(r-r_+) ( r- r_c) ( r+ r_{--}), \quad 0 < r_- \leq r_+ \leq r_c.
\end{eqnarray}
The three positive roots at $r= r_{-}, r_{+}$ and $r_{c}$  correspond to the inner, the outer and the cosmological horizon respectively, while the other negative root at $r=-r_{--} (r_{--}>0)$ is considered unphysical.  

In anti-de Sitter case, the function $\Delta_r(r)$ takes generically the following form;
\begin{equation}
\Delta_r (r)=  \frac{1}{l^2}( r- r_-)(r-r_+) ( r^{2}+p r + q), \quad 0 < r_- \leq r_+ ,
\end{equation}
with $p^{2}-4q<0$. Here, we have at most two horizons.

\FIGURE{
\centering
\includegraphics[width=15 cm]{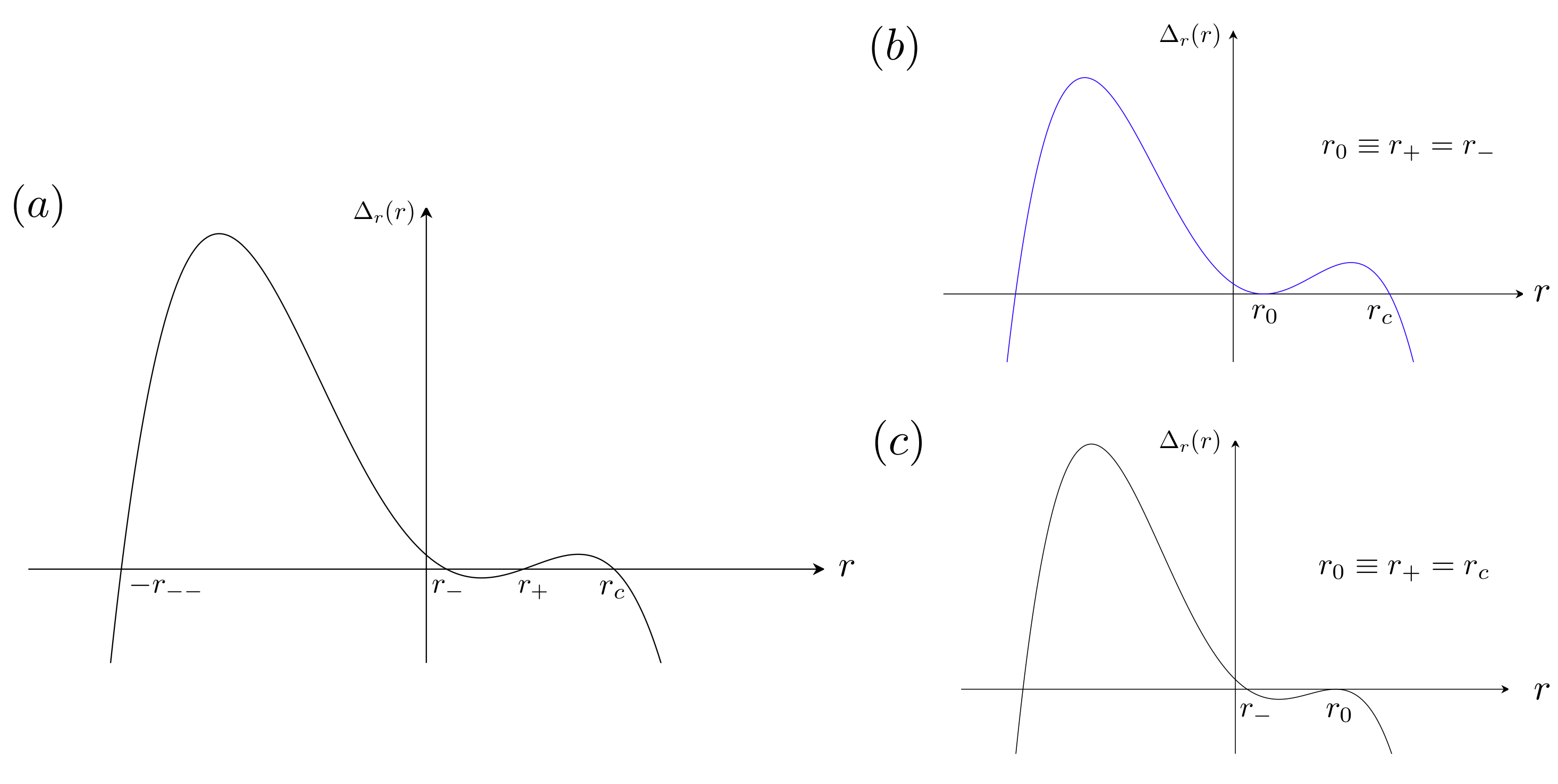}
        \caption{The metric function $\Delta_r(r)$ describing various horizons of $4$-dimensional Kerr-de Sitter black holes. (a) Non-degenerate horizons (b) The horizons of an extremal Kerr-de Sitter black hole (c) The horizons  of a Nariai type black hole.  }
	\label{Gmass1}
}
\FIGURE{
\centering
\includegraphics[width=15 cm]{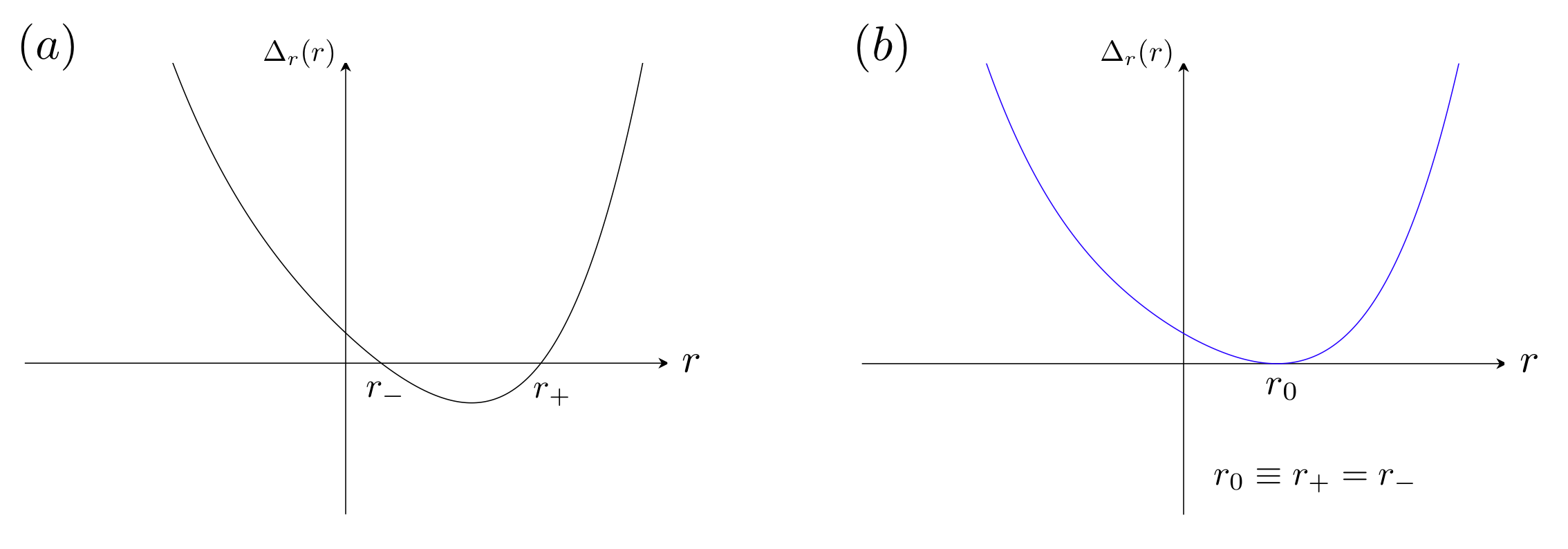}
        \caption{Various horizons of $4$-dimensional Kerr-anti-de Sitter black holes. (a) Non-degenerate horizons (b) The horizons of an extremal Kerr-anti-de Sitter black hole.  }
	\label{Gmass2}
}

\subsection{Extremal Kerr-(anti-)de Sitter Case}

Let us specify the black holes to be discussed hereon.
First, we are interested in the black hole with the degenerate horizons because its geometry is nicely factorized (thus making the entropy function analysis manageable) near those degenerate horizons. In principle, there could be many different cases with the degenerate horizons. For example in Kerr-de Sitter black holes, there are two different cases; the extremal black hole with $r_-$ coincident with $r_+$ (Fig. \ref{Gmass1}(b)), or a Nariai type black hole\footnote{Nariai type black hole is a black hole whose near horizon geometry contains two dimensional de Sitter space-time as defined in  Ref.\cite{Cho:2007we}.} if $r_+$ touches $r_c$ (Fig. \ref{Gmass1}(c)).
We will focus on the former case, that is the extremal black holes. The other case of Nariai type black holes was discussed in  Ref.\cite{Cho:2007we}. We also consider the extremal Kerr-anti-de Sitter black hole, which allows up to one degenerate horizon, since there is no cosmological horizon. (See Fig. \ref{Gmass2}(b).)

In each of these cases, the position of horizons can be expressed in terms of the parameters $m, a$ and $l$. When $\Delta_r(r)$ has a double zero at $r=r_0$, we can factorize the function as follows:
 
\begin{eqnarray}\label{c11}
\Delta_r (r)  &=& (r^2 + a^2 ) \biggl( 1- \frac{\eta}{l^2} r^2 \biggr) - 2 m r \nonumber\\
&=& \left\{ 
\begin{array}{l l}
 -\frac{\displaystyle 1}{\displaystyle l^2}\,( r- r_0)^2 ( r- r_c) ( r+ (2 r_0+r_c)) & \quad \mbox{if $\eta = 1,$}\\
~~\,\frac{\displaystyle 1}{\displaystyle l^2}\,( r- r_0)^2 (r^2 + 2 r_0 r + 3 r_0^2 + a^2 +l^2) & \quad \mbox{if $\eta = -1.$}\\
 \end{array} \right.
\end{eqnarray}
Here, the value $\eta=1$ is for Kerr-de Sitter black holes while $\eta=-1$ is for Kerr-anti-de Sitter black holes. To be more general, let us assume $r_c$ as a positive simple zero of $\Delta_r(r)$ not necessarily to be the cosmological horizon for the moment.
In this factorization, we can find the relations between the horizons and the parameters ($m, a$ and $ l$). We will consider de Sitter case and anti-de Sitter case one by one.
\\
\\
\noindent
(i) de Sitter case ($\eta=1$)

From Eq.(\ref{c11}), we find the relations
\begin{eqnarray}
&&\frac{r_0^4}{l^2} - \biggl(1- \frac{a^2}{l^2}\biggr) r_0^2 + 2 m r_0 - a^2 = 0, \label{h1}\\
&&\frac{4 }{l^2} r_0^3 -2 \biggl(1-\frac{a^2}{l^2} \biggr) r_0 + 2 m =0,\label{h2}\\
&&\frac{1}{l^2} \left(3 r_0^2 + 2 r_0 r_c + r_c^2 \right) - \biggl(1-\frac{a^2}{l^2} \biggr) = 0.\label{h3}
\end{eqnarray}
The extremality heavily restricts the range of $r_0$. Solving  Eq.(\ref{h3}) for $r_c$ in terms of $r_0$, we have 
\begin{eqnarray}\label{root}
r_c = - r_0 \pm \sqrt{-2 r_0^2 +( {l^2} - a^2 )}\,.
\end{eqnarray}
We take the positive value of $r_c$ (the upper sign). For the following range of $r_0$, 
\begin{eqnarray}\label{range}
0\leq r_0^2 <\frac{1}{6}(l^2 - a^2),
\end{eqnarray}
we have $r_c > r_0$, which corresponds to an extremal black hole. The other case of $r_c < r_0$ corresponds to a Nariai type black hole.

\FIGURE{
\centering
\includegraphics[width=15 cm]{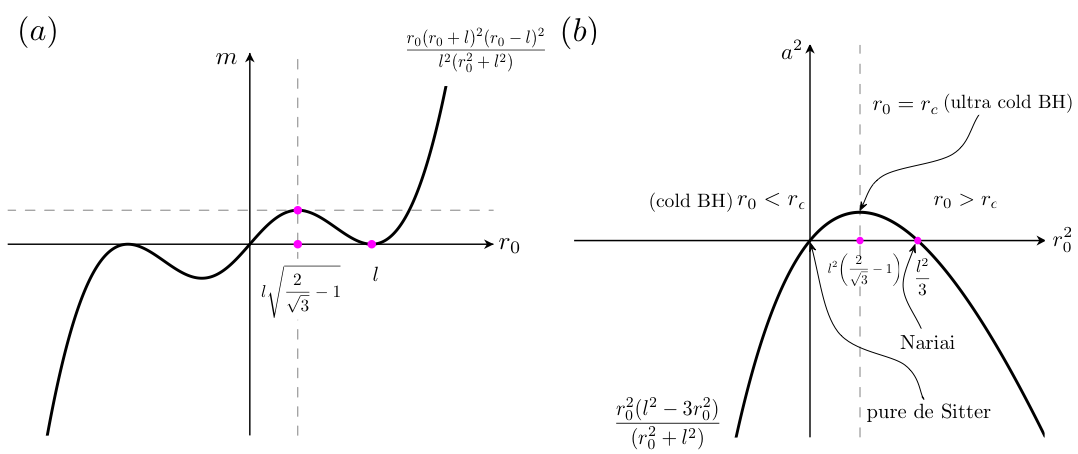}
        \caption{Parameters in Kerr-de Sitter black holes (a) Mass parameter $m$ as a function of the position $r_0$ of the degenerate horizon (b) Angular momentum parameter $a^2$ as a function of the position $r_0^2$ of the degenerate horizon}
	\label{gds}
}

One can go further and write the range of $r_0$ in terms of $l^2$ only. Eqs.(\ref{h1}) and (\ref{h2}) enable us to eliminate $m$, and write $r_0$ in terms of $l$ and $a$;
\begin{eqnarray}\label{c1}
r_0^2 = \frac{l^2}{6 }\,\biggl\{\biggl(1- \frac{a^2}{l^2}\biggr) \pm \sqrt{\biggl(1-\frac{a^2}{l^2} \biggr)^2 - \frac{12 a^2}{l^2}}~\biggr\}.
\end{eqnarray}
The lower sign corresponds to the extremal case of our interest. On the other hand, the upper sign describes a Nariai type black hole.

We observe that the upper bound in Eq.(\ref{range}) is nothing but the point of $r_0^2$ where $a^2$ takes the extremum value. Indeed it is the very point where the square root part of Eq.(\ref{c1}) vanishes. Here we have the relation,
\begin{equation}\label{arange}
a^2 = l^2 \left( 7\pm 4\sqrt { 3}\,\right).
\end{equation}
For the lower sign in this equation, we can rewrite Eq.(\ref{c1}) as follows:
\begin{equation}\label{dsr}
0\leq r_0^2 < l^2\left(\frac{2}{\sqrt{3}} -1 \right).
\end{equation} 
The upper sign of Eq.(\ref{arange}) gives negative $r_0^2$, so neglected.
\FIGURE{
\centering
\includegraphics[width=15 cm]{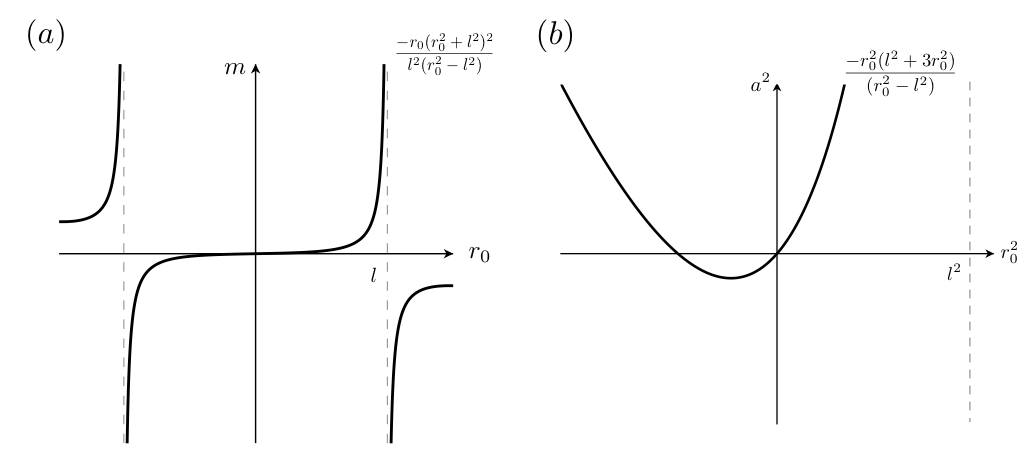}
        \caption{Parameters in Kerr-anti-de Sitter black holes (a) Mass parameter $m$ as a function of the position $r_0$ of the degenerate horizon (b) Angular momentum parameter $a^2$ as a function of the position $r_0^2$ of the degenerate horizon }
	\label{gads}
}
\\
\\
\noindent
(i) anti-de Sitter case ($\eta=-1$)

From Eq.(\ref{c11}), we find the relations
\begin{eqnarray}
&&\frac{r_0^4}{l^2} + \biggl(1+ \frac{a^2}{l^2}\biggr) r_0^2 - 2 m r_0 + a^2 = 0, \label{hh1}\\
&&\frac{4 }{l^2} r_0^3 +2 \biggl(1+\frac{a^2}{l^2} \biggr) r_0 - 2 m =0.\label{hh2}
\end{eqnarray}
In this case, there is no bound coming from the `cosmological' horizon. However, there still is a restriction on the range of $r_0$. First of all, $r_0$ has to be positive. Fig. {\ref{gads}} shows that we must have  $0 \leq r_0 < l$ in order to have $m$ and $a^2$ positive. 
Also, to satisfy $a^2 < l^2$ assumed in Sec. 2.2, we must have $0\leq r_0^2 < {l^2 \over 3} $. Therefore, we can conclude that range of $r_0$ for the extremal Kerr-anti-de Sitter black hole is
\begin{eqnarray}\label{adsr}
0\leq r_0^2 < \frac{l^2}{3}.
\end{eqnarray} 

From Eqs.(\ref{hh1}) and (\ref{hh2}) the horizon size of the extremal Kerr-anti-de Sitter black hole is proportional to 
\begin{eqnarray}\label{cc1}
r_0^2 =-{{l^2}\over {6 }} \,\biggl\{\biggl(1+ {{a^2}\over l^2}   \biggr) \pm \sqrt{\biggl(1+{{a^2}\over {l^2}} \biggr)^2 + {{12 a^2 }\over {l^2}} }~\biggr\}.
\end{eqnarray}
Only the lower sign gives physical $r_0^2$.

\subsection{Hawking Temperatures}
Let us consider the Hawking temperatures of the extremal Kerr-(anti-)de Sitter black holes. One simple way of computing Hawking temperature is to use the Euclidean method \cite{Gibbons:1976ue}. It is given by\begin{equation}\label{temp}
T = {{|\Delta_r'(r_H)|}\over{4 \pi (r_H^2 + a^2)}},
\end{equation}
where $r_H$ stands for zeros of $\Delta_r(r)$.

So the extremal black hole has zero temperature. For the extremal Kerr-de Sitter case, we have two horizons.  Each horizon has different temperature. The event horizon (at $r_0$) has zero temperature. However, there is non-vanishing temperature of the cosmological horizon,
\begin{equation}
T_{r_0}= 0, \quad T_{r_c} = {{(r_c - r_0)^2 (r_c +r_0 )}\over{2 \pi l^2 (r_c^2+a^2)}}.
\end{equation}
We call this `cold' black hole. Of course, in the special case of $r_0=r_c$, $T_{r_c}$ is also zero, and this case is called an `ultra cold black hole'  \cite{Cho:2007mn}.

\subsection{Bekenstein-Hawking Entropy}
At the event horizon of the 4-dimensional extremal Kerr-(anti-)de Sitter black hole, the Bekenstein-Hawking Entropy is given as
\begin{eqnarray}\label{bhe}
S &=& {A\over{4}}= {{\pi l^2 ( r_0^2+a^2)}\over{l^2 + \eta a^2}}.
\end{eqnarray}
Eliminating $a$ by using a relation Eq.(\ref{c1}) or Eq.(\ref{cc1}), the above result can be rewritten as \begin{equation}\label{eq32}
S={{2
    \pi l^2 r_0^2 }\over{\, l^2 + 3 \eta r_0^2\,  }}.
\end{equation}
The entropy of an extremal rotating black hole is determined by angular momentum of the black hole thanks to the relation between mass and angular momentum. Using Eqs.(\ref{mj}) and (\ref{c1}) (or (\ref{cc1})), one can write the angular momentum as
\begin{eqnarray}\label{j1}
J &=& {{a \,( r_0^2 + a^2 )}\over {2 r_0}} \biggl({1-{ \eta \over
l^2 }r_0^2}\biggr)\biggl( 1+ {\eta \over l^2}a^2 \biggr)^{\!-2}\! = {{r_0^2\, l^2 \sqrt{(l^2 - 3 \eta r_0^2\,) ( l^2 + \eta r_0^2
  \,)}}\over{(l^2+3 \eta r_0^2 \,)^2}}.~~~
\end{eqnarray}
We will obtain an algebraic relation between the entropy and the angular momentum. We can invert  Eq.(\ref{eq32}) and get
\begin{eqnarray}
r_0^2 = {{  l^2 S}\over{2 \pi l^2 - 3\eta S}}.
\end{eqnarray}
Inserting this into Eq.(\ref{j1}) results in 
\begin{equation}\label{r1}
3 S^4 - 4 \eta \lambda^2 S^3 + \lambda^4 S^2 - 4 \pi^2 \lambda^4 J^2 =
0, 
\end{equation}
where $\lambda^2= \pi l^2$. Given an angular momentum, one can determine the entropy. In principle, one can solve this equation for the entropy. However, we will not give the messy expression of the entropy here because the above relation is sufficient for our purpose. At the final stage, we will compare the above relation with the one obtained by the entropy function formalism.

\subsection{Isometry of the Near Horizon Geometry}
Let us consider a near horizon geometry of the extremal Kerr-(anti-)de Sitter black holes. 
In the near horizon coordinates $(\bar{t},\,\bar{r},\,\bar{\phi})$ defined as
\begin{equation}
{ r}=r_0 + \epsilon {\bar r}, ~~~ { t} = {{\bar t} \over \epsilon},~~~{ \phi} = {\bar\phi}+{{a \Xi {\bar t}}\over
  {(r_0^2 + a^2 ) \epsilon}},
\end{equation}
 the near horizon geometry is obtained by taking a limit $\epsilon \rightarrow 0$. Since $\Delta_r(r)$ has a double zero for the extremal case, its expansion starts from the second order in $\epsilon$;
\begin{eqnarray}
\Delta_r (r) ={{\Delta_r'' (r_0)}\over 2} (\epsilon {\bar r})^2+ {\cal{O}} (\epsilon^3).
\end{eqnarray}
Then the metric of Eq.(\ref{kds1}) takes the following form; 
\begin{eqnarray}
ds^2&=& {{\rho_0^2}\over{r_0^2 +a^2}} \biggl(-{{\Delta''(r_0)}\over{2 (r_0^2+a^2)}}~{\bar r}^2 d{\bar t}^{\,2} +
  {{2 (r_0^2+a^2)}\over{ \Delta''(r_0)}}~{d{\bar r}^2\over {\bar r}^2}~\biggr) \nonumber\\ 
  &&+ {{\rho_0^2}\over{\Delta_{\theta}}} ~d \theta^2 +{{\Delta_{\theta}
      ~(a^2+r_0^2)^2 \sin^2 \theta}\over{\Xi^2 \rho_0^2}}\biggl(d{\bar \phi}+{{2 a r_0
      \Xi}\over{(a^2+r_0^2)^2}}\,{\bar r} d{\bar t} ~\biggr)^2, \label{nhg1}
\end{eqnarray}
where
\begin{eqnarray}
\Delta_r'' (r_0) = 
2 \biggl(1-{\eta \over l^2} \left(6 r_0^2 +a^2\right)\biggr),\quad\rho_0^2 = r_0^2 + a^2 \cos^2 \theta .
\end{eqnarray}
Though in the case of the extremal black hole of which the horizon size $r_0$ and the angular momentum parameter $a$ are related to each other by Eq.(\ref{c1}), we keep both parameters, for the moment, to make expressions simpler. 

Introducing a rescaled radial coordinate ${\tilde{ r}}$ as
\begin{equation}
{\tilde r} = {{\Delta''(r_0)}\over{2(a^2 + r_0^2)}}~{\bar r},
\end{equation}
we can simplify the metric Eq.(\ref{nhg1}) as
\begin{eqnarray}\label{nhg2}
ds^2 &=& {{2 \rho_0^2}\over{\Delta''(r_0)}}\biggl(\!- {\tilde r}^2 d{\bar t}^2 +
  {d{\tilde r}^2\over {\tilde r}^2}\biggr)\! + \!{{\rho_0^2}\over{\Delta_{\theta}}} d \theta^2\!+\!{{\Delta_{\theta}
      (r_0^2+a^2)^2 \sin^2 \theta}\over{\Xi^2 \rho_0^2}}\biggl(\!d{\bar\phi}+{{ a r_0
      \Xi\, {\tilde r} d{\bar t}}\over{(a^2+r_0^2) \Delta''(r_0)}} \!\biggr)^2\!.\nonumber\\
      &&
\end{eqnarray}
It clearly shows that the near horizon geometry around the double zero $r_0$ is AdS$_2 \times$ S$^1$ with the isometry $SO(1,2) \times U(1)$.

\section{Entropy Function of Kerr-(anti-)de Sitter Black Holes}

In this section, we will demonstrate that Sen's entropy function method works for Kerr-(anti-)de Sitter black holes. We  recover the result of Eq.(\ref{r1}) without the knowledge of the exact form of the solution of the Einstein equation. In Sen's entropy function method, all we need for the computation of the entropy of an extremal black hole is its near horizon geometry. In the static extremal cases, the near horizon geometry, respecting the assumed isometry of $SO(1,2)\times SO(3)$, is characterized by some parameters depending on the mass, the charge, and the angular momentum of the black holes. The near horizon geometry of the stationary black holes is more complicated than the static cases. The above parameters are generalized to some unknown functions like $f(\theta),\,g(\theta),$ and $h(\theta)$, which depend not only on the mass, the charge, and the angular momentum, but also on the latitudinal angle $\theta$. 

In order to carry out Sen's entropy function method for the extremal Kerr-(anti-)de Sitter black holes, we first make an ansatz for their near horizon geometries. The unknown functions $f(\theta), g(\theta)$ and $h(\theta)$ appearing in the ansatz are governed by the `Euler-Lagrange equations' extremizing the Lagrangian. We then solve the Euler-Lagrange equations with proper boundary conditions, and determine these functions. Inserting the results into the entropy function, we obtain the black hole entropy, which coincides with the Bekenstein-Hawking entropy. Finally we also consider the correction of the black hole entropy coming from the Gauss-Bonnet term in the Lagrangian.

\subsection{An Ansatz for the Metric}
Wald's entropy formula \cite{{Wald:1993nt},{Jacobson:1993vj},{Iyer:1994ys}}, applied to the  near horizon geometry of an extremal black hole, results in the entropy function. It takes the form of some Legendre transformation of Lagrangian density (over AdS$_2$ spacetime) with respect to some parameters characterizing the near horizon geometry \cite{{Sen:2005wa},{Sen:2005iz}}. This method can be generalized to the stationary ones with $SO(1,2) \times U(1)$ near horizon isometry \cite{Astefanesei:2006dd} or to the static cases in de Sitter background \cite{Cho:2007mn}.
We will verify that the method is also successful for the case of the stationary black holes in the background with cosmological constant. 

Let us start with the action for the gravity theory in (anti-)de Sitter background,
\begin{equation}
{\cal{S}} = {1 \over {16 \pi}}\int d^4 x \sqrt{-g}~ ( R - 2 \Lambda_4 ).
\end{equation}
Based on the near horizon isometry $SO(1,2) \times U(1)$ of extremal stationary black holes, we make an ansatz for the near horizon geometry as 
\begin{eqnarray}\label{a1}
ds^2 = f^2(\theta)~(-r^2 dt^2 + {dr^2 \over r^2}) + g^2(\theta)~d \theta^2 + h^2(\theta)
~{\rm{sin}}^2 \theta ~(d\phi + \alpha\, r\, dt)^2.
\end{eqnarray}
Here $\alpha$ is a dimensionless constant and $f, g, h$ (with the length dimension) are functions of $\theta$. The latitudinal angle $\theta$ takes values $0 \leq \theta \leq \pi$ while the azimuthal angle $\phi$ is a periodic coordinate with period $2\pi$. 

We give some physical conditions on the metric coefficients, $ f^2(\theta),~ g^2(\theta)$, and $h^2(\theta)$.
Since the geometry is symmetric across the equator, we first require that 
\begin{eqnarray}\label{cond1}
&&f^2(\theta)=f^2(\pi-\theta),\quad g^2(\theta)=g^2(\pi-\theta),\quad h^2(\theta)=h^2(\pi-\theta).
\end{eqnarray}
Second, in order to make the metric, in $\Lambda_{4}\rightarrow 0$ limit, approach the near-horizon metric of the extremal Kerr black hole, the functions, $ f^2(\theta),~ g^2(\theta)$, and $h^2(\theta)$ should be regular and also be non-vanishing. 
Lastly, we give the following condition
\begin{equation}\label{bc5}
 g(\theta) h(\theta) = k,
 \end{equation}
 where $k$ is a nonzero constant carrying the dimension of length squared. This condition means that we keep the volume of the sphere part constant as it is deformed by the rotation. It will greatly simplify the equations of motion which we will solve later. We will see this in the next subsection. Furthermore, this accords with the property of Kerr metric if we let $\Lambda_{4}\rightarrow 0$.

Let us comment on the boundary conditions. Since the rotation does not distort the geometry  near the north-pole and the south-pole, we have
  $g^2(\theta)=h^2(\theta)$ at $\theta= 0, \pi$.
Combined with the condition of Eq.(\ref{bc5}), it leads to;
\begin{equation}\label{bc4}
g^2(\theta)=h^2(\theta)= k, \quad{\rm{at}}~~ \theta= 0, \pi.
\end{equation}
Now we have three unknowns $f(\theta)$, $h(\theta)$ and $\alpha$ to be  determined. 

\subsection{Euler-Lagrange Equations} 
Given the cosmological constant $\Lambda_{4}$, and angular momentum $J$ which is conjugate to the parameter $\alpha$, we will determine the forms of $\alpha$, $f(\theta)$, and $h(\theta)$ so that they extremize the entropy function.

In terms of the angular momentum $J$, the entropy function, being defined as the Routhian, takes the form
\begin{equation}
F = 2 \pi ( J \alpha -  L ).
\end{equation}
Here, the function $L$ is the Lagrangian density over two-dimensional
anti-de Sitter spacetime. We write down its explicit expressions and the equations of motion below.
 \begin{eqnarray}\label{efp}
L &=& {1\over{16 \pi}} \int d\theta d \phi \sqrt{-g} ~\cal{L}\nonumber\\
&=& {1 \over 4}\int d \theta ~{{\sin \theta}}  \biggl\{ \left( f^2\,h+f'^2\,h +2 f f' h' \right) {1\over g} + \left( g' h - g h' \right) {{f^2 \cot \theta}\over {g^2}} - g h - {{3 \eta f^2 g h }\over l^2}      \biggr\} \nonumber\\
&&-{{1}\over{4}}  {\biggl[ \left( 2 f f' h \sin\theta + f^2 h' \sin\theta \right) {1\over g} \,\biggr]}_{\theta =
0}^{\theta=\pi} 
\end{eqnarray}
The boundary terms arise from integration by parts. In the above, for convenience, we used simplified notations for functions of $\theta$ such as $f$ and $ h$ instead of $f(\theta)$ and $h(\theta)$, and  the notation $'$ for the derivative with respect to the coordinate $\theta$.  The black hole entropy is given by the extremum value of the entropy function $F$.  As the first step to obtain this, we follow the standard variational procedure which results in the Euler-Lagrange equations as
\begin{eqnarray}
{{\delta F}\over {\delta f(\theta)}} = 0\, :~
&&\left(f ' g h \right)'  - 2 f' g' h + ( 2 f g h' - f g' h + f' g h) \cot \theta \nonumber\\
&&~ + \left( g'h' - gh - g h''\,\right) f + {{\alpha^2 g^3 h^3 \sin^3 \theta}\over{4 f^3}} + {{3 \eta f g^3 h}\over l^2} = 0,\label{ep1}\\
{{\delta F}\over {\delta  g(\theta)}} =0\,:~~
&&g^2 +f'^2 + {{3 \eta f^2 g^2 }\over{l^2}} + 2 f f'' - {{2 f f' g' }\over g} - {{3 \alpha^2 g^2 h^2 \sin^2 \theta }\over {4 f^2}} = 0,\label{ep2}\\
{{\delta F}\over {\delta    h(\theta)}} =0\,: ~~
&&g^2 +f'^2 + {{3 \eta f^2 g^2 }\over{l^2}} + 2 f f' \cot \theta + {{2 f f' h' }\over h} - {{ \alpha^2 g^2 h^2 \sin^2 \theta }\over {4 f^2}} =0 \label{ep3},~\qquad\\
{{\partial F}\over {\partial \alpha}} = J \,: ~~&& {{\alpha}\over 8} \int d\theta~ {{ g h^3 \sin^3 \theta}\over f^2} = J.
\end{eqnarray}
Notice that when we combine Eq.(\ref{ep2}) and Eq.(\ref{ep3}), we have rather a simple differential equation,
\begin{eqnarray}\label{}
f'' - f' \left[ \left(\ln ({gh}) \right)' + \cot \theta \right] - {{ \alpha^2 \left(g  h\right)^2 \sin^2 \theta }\over {4 f^2}}=0.
\end{eqnarray}
This equation completely decouples from $g(\theta)$ and $h(\theta)$, if we use the condition $g(\theta) h(\theta) = k$. This condition also simplifies Eq.(\ref{ep1}). One can justify the validity of the condition by checking whether the results of the unknowns thus obtained indeed solve the equations of motion.  We have checked that this is the case. 

Let us now give the simpler version of the expressions incorporating the condition $g(\theta) h(\theta) = k$ as follows:
\begin{eqnarray}\label{ef}
L &=& {1\over{16 \pi}} \int d\theta d \phi \sqrt{-g} ~\cal{L}\nonumber\\
&=& {1 \over 4}\int d \theta ~{{\sin \theta}\over k}  \biggl\{ \biggl(f^2+
f'^2+ { { \alpha^2 k^2  
\sin^2 \theta}\over{4 f^2}}\biggr)p+ (f f' -f^2 \cot \theta)p' -k^2- {{3 \eta k^2 f^2
}\over{l^2}}\biggr\} \nonumber\\
&&-{{1}\over{8 k}}  {\biggl[4 f f' p \sin\theta
+ f^2 p'  \sin\theta \biggr]}_{\theta =
0~,}^{\theta=\pi} 
\end{eqnarray}
where $p(\theta) \equiv h^2(\theta)$. Now we have three unknowns, $f(\theta), p(\theta)$ and $\alpha$, satisfying the following equations of motion:
\begin{eqnarray}
p'' +  \biggl({{2 f'}\over{f}} +3 \cot\theta \biggr)
&p'&+ \biggl( {{\alpha^2 k^2 \sin^2
 \theta}\over{2 f^4}}+ {{2( f'' +
 f'\cot\theta)} \over {f}} -2 \biggr) p+{{ 6\eta  k^2}\over{l^2}} =0 ,\cr
 &&\label{e1}\\
&&f'' - f' \cot \theta - {{\alpha^2 k^2 \sin^2 \theta}\over{4 f^3}} =0 ,\label{e2}\\
&& {{\alpha k }\over 8}\int d \theta ~ {{ p~ \sin^3 \theta}\over{f^2}}=J.\label{e3}
\end{eqnarray}
The variations for the boundary terms are
\begin{eqnarray}\label{vb}
\delta F = {\pi \over{ 4 k}} \biggl[\, 2(\,f^2 \cos\, \theta + f f' \sin\, \theta \,) \,\delta p + f^2 \,\sin^2 \theta\, \delta p' + 4 fp  \,\sin \,\theta \,\delta f' \biggr]_{\theta=0}^{\theta=\pi}.
\end{eqnarray}
It vanishes due to the conditions in Eq.(\ref{cond1}). 

Eq.(\ref{e2}), written purely in terms of $f(\theta)$, is easy to solve despite its nonlinearity.  
In terms of $x\equiv\cos \theta$ and $\beta^2 \equiv \alpha^2 k^2/4$,  the equation gets  simplified as
\begin{eqnarray}\label{d1}
{{d^2 f(x)}\over {dx^2}} - {{\beta^2}} f^{-3} (x) = 0.
\end{eqnarray}
This allows us to determine the form of $f^2(\theta)$, up to two integration constants $b$ and $c$, as  
\begin{equation}\label{ff}
f^2(\theta)= {{{\beta^2}\over{ b}} + b \,(\cos \theta -c)^2}.
\end{equation}
Here, the constant $b$ should be positive and carries the dimension of the length squared. The other constant $c$ vanishes because of Eq.(\ref{cond1}).

Solving Eq.(\ref{e1}) is a bit tricky. For this, it is very important to note that the equation is nothing but the $\theta\theta$-component of the Einstein equation $R_{\theta\theta}-\Lambda_{4}g_{\theta\theta}=0$ applied for the metric ansatz (\ref{a1}). (In order to show this, one needs to use Eq. (\ref{e2}).)  This is clear in the explicit form of the component $R_{\theta \theta}$ for the geometry (\ref{a1}):
\begin{equation}
R_{\theta \theta} = {1 \over {2 p}}\biggl\{ \biggl( 2 - {{4 f''}\over f}\biggr) p - \biggl({{2 f'}\over f} + 3 \cot \theta \biggr) p' - p'' \biggr\}.
\end{equation}
As it is, the expression is a second order differential equation for $p(\theta)$.  
One can rewrite this into a first order differential equation, which is much easier to solve.  
The trick is based on the fact that the equation, $R_{\theta\theta}-\Lambda_{4}g_{\theta\theta}=0$, is equivalent to $G_{\theta\theta} +\Lambda_4 g_{\theta\theta} = 0$, for the Einstein tensor $G_{\mu\nu} = R_{\mu\nu}- {1\over2} R g_{\mu\nu}$. Since 

\begin{equation}
G_{\theta \theta} =  {{f'} \over {f  p}}\biggl\{ p' +  \biggl( {{f'}\over f}- {{\beta^2
    \sin^2 \theta}\over{ f^3 f'}} + 2 \cot \theta \biggr) p+ {{k^2}\over{f f'}}  \biggr\},
\end{equation}
we have 
\begin{eqnarray}\label{d2}
G_{\theta\theta} +\Lambda_4 g_{\theta\theta}= p' +  \biggl( {{f'}\over f}- {{\beta^2   \sin^2 \theta}\over{ f^3 f'}} + 2 \cot \theta \biggr) p + {{k^2}\over{f
    f'}}+{{ 3\eta k^2  f}\over {l^2 f'}} = 0,
\end{eqnarray}
which  is a first order differential equation. Since, $f(\theta)$ was already calculated in Eq.(\ref{ff}), the explicit form for $p(\theta)$ can be obtained, as we will see in the next subsection.

\subsection{Exact Form of the Near Horizon Metric}

The exact solution of Eq.(\ref{d2}) is
\begin{eqnarray}\label{sol1}
p(\theta) \!&=& \!{{k^{2}}\over{({{\beta^2}}+b^2\cos^{2}\! \theta ) \sin^2 \!\theta}}\biggl\{ {{\beta^2}\over { b  }} -\! b \cos^2\!\theta - {{n \cos \theta}\over{ k^2}} + \!{{3\eta }\over {l^2}}\biggl( {{\beta^4}\over { b^2 }} -  {{2 \beta^2}}\!\cos^2\!\theta  - {{b^2 \cos^4\!\theta}\over 3}\biggr)\! \biggr\}\!, \nonumber\\
&&
\end{eqnarray}
where $n$ is another integration constant which should vanish owing to the condition of Eq.(\ref{cond1}). 

One can remove other constants $\beta$ and $k$ further so that only the constant $b$ remain in the final results of the unknowns. For this, we use first the condition that the function be regular at $\theta=0, \pi$. To cancel the possible singularity at $\theta = 0, \pi$ due to $\sin^2 \theta$ factor, the numerator must vanish at $\theta = 0, \pi$. This results in the relation among the constants;  
\begin{eqnarray}\label{bbk}
3 \beta^4 +  b ( \eta l^2 - 6 b ) \beta^2 - b^3 ( b + \eta l^2) &=& 0.
\end{eqnarray}
This allows us to write $\beta$ in terms of $b$;
\begin{eqnarray}\label{beta}
\beta^2 = {{ b} \over 6} \,\left( \, 6 b - {\eta\, l^2 } \pm \sqrt{l^4 + 48 b^2} \,\right)
\end{eqnarray}
and the function $p(\theta)$ gets simplified as
\begin{equation}\label{pth}
p(\theta) = {k^2\over {~{\beta^2} + b^{2} \cos^2 \theta}} \biggl( {{\beta^2 ( b\, l^2 + 3\eta \beta^2 )}\over { b^2 l^2}} + {{\eta\, b^{2}}\over {l^2}} \cos^2 \theta \biggr).
\end{equation}

The choice of multi-sign in Eq.(\ref{beta}) depends on the sign of the cosmological constant; the {\it upper sign} for de Sitter and the {\it lower sign} for anti-de Sitter case. The choice ensures the positivity of $\beta^2$ and $p(\theta) ( \equiv h^2 (\theta))$. We defer the details to the appendix \ref{apb1}. There, we also find that anti-de Sitter case restricts the range of $b$ as $0< b<{{l^2}/{4}}$.

Having these conditions in mind, we can rewrite Eq.(\ref{beta}) for  both de Sitter and anti-de Sitter cases at once using an unified notation inserting $\eta$ as
\begin{eqnarray}
\beta^2 = {{ b} \over 6}\, \left(\,  6 b - {\eta \,l^2}  + \eta \sqrt{l^4 + 48 b^2}\,\right).
\end{eqnarray}
So far, we have obtained form of $f(\theta)$ and $p(\theta)$ up to unknown constants $k$ and $b$. From the boundary condition (\ref{bc4}),  $k$ can be written in terms of $b$ as;
\begin{eqnarray}
k &=& ({\beta^2}+{ b^2})\biggl({{\beta^2( b l^2 + 3 \eta \beta^2 )}\over { l^2 b^2}} + {{\eta b^2} \over l^2}\biggr)^{-1}= {{l^2 \left( 12 b + \eta l^2 - \eta\sqrt{l^4 + 48 b^2}\,\right) }\over {6~ (l^2+4 \eta b)}}.~~
\end{eqnarray}
Therefore, the unknowns $f(\theta), p(\theta)$ and $\alpha$ are written only in terms of $b$ as follows: 
\begin{eqnarray}\label{rb}
f^2 (\theta)&=& {1\over 6} \left( 6 b - \eta l^2 + \eta\sqrt{l^4 + 48 b^2}\right) + b \cos^2 \theta ,\nonumber\\
p(\theta) &=& {{l^2 \left( 12 b + \eta l^2 - \eta\sqrt{l^4 + 48 b^2}\,\right)^2 }\over {6 ~(4 \eta b + l^2)^2}}\biggl({{7 b + \eta\sqrt{l^4 + 48 b^2} + b \cos^2 \theta} \over{ 6 \eta b - l^2 + \sqrt{l^4 + 48 b^2}+ 6\eta b \cos^2 \theta}}\biggr),\nonumber\\
\alpha&=& {{2 \beta} \over k}= {{\left( 8 \eta b +\eta l^2 \sqrt{l^4 + 48 b^2} \,\right)\left( 6 b - \eta l^2 +\eta\sqrt{l^4 + 48 b^2}\,\right)^{1/2} }\over { \sqrt{24 b}~  l^2 }} .
\end{eqnarray}
The interpretation of the constant $b$ is unclear in the case at hand, but we will see later that it corresponds to the angular momentum $J$ in the limit of zero cosmological constant.

\subsection{Extremum Value of the Entropy Function}

Let us remind ourselves that we set $p(\theta)=h^2(\theta)$ and $\beta^2 = \alpha^2 k^2/4$ above and only one undetermined constant $b$ is remained to be fixed. Now we need to consider Eq.(\ref{e3}), the last Euler-Lagrange equation to be solved. Inserting Eq.(\ref{rb}) into Eq.(\ref{e3}) we get a relation between the angular momentum $J$ and the constant $b$ as
\begin{eqnarray}\label{jb}
J &=& {\beta \over 8}\int d \theta~ {{ p (\theta) \sin^3 \theta}\over {f^2(\theta)}}\nonumber\\
&=& {{l^2 \sqrt{b}~\left(l^2 - 12\eta b + 2 \sqrt{l^4 + 48 b^2}\,\right)  \left( 6 b -\eta l^2 + \eta\sqrt{ l^4 + 48 b^2}\,\right)^{1/2}}\over{ 3 \sqrt{6}~  \left( 4\eta b + l^2 \right)^2}}.
\end{eqnarray}
From this relation, we can obtain $b$ written in terms of $J$ although the form is very complicated. We will not write down here the exact form of $b$, which is one-page long, but keep the above relation instead. 
Eq.(\ref{rb}) also can be written in terms of angular momentum $J$ eventually. It means that all unknowns we introduced in our ansatz are determined with a given angular momentum $J$. 

Now, we note that the Eq.(\ref{rb}) coincides with the metric coefficients of the near horizon geometry in Eq.(\ref{nhg2}) if we replace $b$ with 
\begin{eqnarray}
b = {{l^2  r_0^2\,( l^2 - 3 \,\eta\, r_0^2)}\over{l^4 - 6\, \eta\, l^2 r_0^2 - 3\,  r_0^4}}\,.
\end{eqnarray}
Deriving this result, we used the extremal conditions of Eq.(\ref{c1}) and Eq.(\ref{cc1}). 
Furthermore, we find that the ranges of $b$, (for de Sitter case, $b>0$ and for anti-de Sitter case, $0< b< {l^2 /4}$), 
match exactly those of $r_0^2$\,. See appendix \ref{apb2} for details.

Now, we will obtain a relation between the angular momentum $J$ and the extremal value of the entropy function $F$ of Eq.(\ref{ef});   
\begin{eqnarray}\label{entftn}
F= {{ \pi l^2 \left( 12\,\eta\, b + l^2 -\sqrt{l^4 + 48 b^2}\,\right) }\over{ 6\, (\, 4 b +\eta\, l^2 \,)}}.
\end{eqnarray}
By inverting the equation, we can write $b$ in terms of $F$ as
\begin{eqnarray}
b = {{ \lambda^2 ( \lambda^2  - 3\eta F)F }\over{2 \pi( 6 F^2 - 6 \eta \lambda^2 F^2 + \lambda^4)}}.
\end{eqnarray}
Inserting this result into Eq.(\ref{jb}), we obtain an algebraic relation between $F$ and $J$;
\begin{equation}\label{algerel}
3 F^4 - 4 \eta \lambda^2 F^3 + \lambda^4 F^2 - 4 \pi^2 \lambda^4 J^2 = 0,
\end{equation}
where $\lambda^2= \pi l^2$. This relation has exactly the same form with Eq.(\ref{r1}). Since it is an algebraic equation, we can conclude that $F=S$, that is to say,  $F$ coincides with the Bekenstein-Hawking entropy for the extremal Kerr-(anti-)de Sitter black holes. 

The results coincide with those of the extremal Kerr black holes if we let $l \rightarrow \infty$. In the limit, we get
\begin{eqnarray}\label{rkerr}
&&f^2 (\theta) \rightarrow b ( 1 + \cos^2 \theta), \quad h^2 (\theta) \rightarrow {{4 b}\over{(1+\cos^2 \theta)}}, \nonumber\\
&&k\rightarrow 2 b ,\quad \alpha = 1 ,\quad J \rightarrow b, \quad S \rightarrow 2 \pi b,
\end{eqnarray}
which are nothing but those of the extremal Kerr solution. Replacing $b$ with $r_0^2$, the above results are identified with the ones in the near horizon geometry of the Kerr black holes of Eq.(\ref{nhkerr}). Eq.(\ref{rkerr}) agrees with the results of Eq.(4.30) in Ref.\cite{Astefanesei:2006dd} if we set $b=J/16 \pi$. (Newton's constant $G_N=1/{16\pi}$ was used in  Ref.\cite{Astefanesei:2006dd}.)

\subsection{Effect of the Gauss-Bonnet Term on the Entropy}

In this section, we will consider the higher order corrections to the entropy of Kerr-(anti-)de Sitter black holes. We first note that the near horizon ansatz of Eq.(\ref{a1}) works for the case involving higher derivative terms in the Lagrangian\cite{Kunduri:2007vf}. However, as for the generic higher order terms, the variation of the entropy function leads us to higher order differential equation, which is difficult to solve. We will consider Gauss-Bonnet combination of the terms\cite{Zwiebach:1985uq}, which, being topological in $4$-dimensions, does not affect the equations, so that one can solve it.

The contribution of the Gauss-Bonnet term to the action is
\begin{eqnarray}\label{sgb}
\Delta{\cal{S}} &=& {1\over{16 \pi}} \int d^4 x \sqrt{-g} {\cal{L}}_{GB} \nonumber\\
&=& {{\xi}\over{16 \pi}} \int d^4 x \sqrt{-g}~(R^{MNPQ}R_{MNPQ}-4 R^{MN}R_{MN} + R^2).
\end{eqnarray}
Here, $\xi$ is a constant whose dimension is length squared. Using the metric ansatz of Eq.(\ref{a1}), ${\cal L}_{GB}$ is obtained  as  
\begin{eqnarray}\label{lgb}
{\cal{L}}_{GB} = {{4 \xi}\over {k^4 f^2 \sin \theta}}&\biggl[& {{ \alpha^2 k^2 p^2 \sin^2 \theta}\over {f^3}}\,  \biggl(  f' \sin \theta - {3\over 2} \,f  \cos \theta \biggr)+p p' \, \biggl(  f'^2 \sin \theta- {{ 3 \alpha^2 k^2 \sin^3 \theta}\over {4 f^2}}  \biggr) \nonumber\\
  &&+~k^2 p' \sin \theta  + 2  f'^2 \,p^2 \cos \theta + 2 \,k^2 p \cos \theta~\biggr]'.
\end{eqnarray}
Inserting this result into Eq.(\ref{sgb}), we get the Lagrangian density over AdS$_2$;  
\begin{eqnarray}
\Delta L =  {{ \xi}\over {2 k^3 }}&\biggl[& {{ \alpha^2 k^2 p^2 \sin^2 \theta}\over {f^3}}\,  \biggl(  f' \sin \theta - {3\over 2} \,f  \cos \theta \biggr)+p p' \, \biggl(  f'^2 \sin \theta- {{ 3 \alpha^2 k^2 \sin^3 \theta}\over {4 f^2}}  \biggr) \nonumber\\
  &&+~k^2 p' \sin \theta  + 2  f'^2 \,p^2 \cos \theta + 2 \,k^2 p \cos \theta~\biggr]_{\theta=0}^{\theta=\pi}~.
\end{eqnarray}

Since it contributes only to the boundary terms, it does not change the
equations of motion of Eqs.(\ref{e1})-(\ref{e3}) and therefore the solutions. However, it
has a non-vanishing boundary value which makes the correction to the extremum
value of the entropy function as 
\begin{equation}\label{shift}
\Delta F = -{{2 \pi \xi}\over k}\biggl[~p(\theta)~ \cos \theta ~\biggr]_{\theta=0}^{\theta=\pi} = 4 \pi \xi.
\end{equation}
The total entropy of the extremal Kerr-(anti-)de Sitter black holes
with Gauss-Bonnet correction terms becomes 
\begin{eqnarray}
S_{\rm{tot}} &=& S + \Delta F \nonumber\\
&=& {{  \pi l^2 \left( 12 b + l^2 - \sqrt{l^4 + 48 b^2}\,\right) }\over{ 6 ( 4 b +\eta l^2 )}}+4 \pi \xi\,.
\end{eqnarray}
This results in a constant addition to the entropy, with its value independent of any charges characterizing the black holes. As was discussed in Ref.\cite{Cho:2007mn} one can understand the entropy as a relative quantity so that the constant addition is nothing but a shift of the reference point of the black hole entropy. 

It is very interesting to see that the value of the constant addition to the entropy is the same as the ones for Reissner-Nordstr\"{o}m-(anti-)de Sitter black holes\cite{Cho:2007mn} and for the black holes of Nariai class\cite{Cho:2007we}. We will see later that this universal contribution of Gauss-Bonnet term to the black hole entropy comes as the result of the topological nature of Gauss-Bonnet term in $4$-dimensions. 

Though the parameter $\xi$ introduced in this paper is an arbitrary free parameter, there is an argument determining its value\cite{Aros:1999id,Olea:2005gb}. If we require the whole action including the Gauss-Bonnet term be extremal for the class of the asymptotically (anti-)de Sitter geometries, it should be related to the cosmological constant as
\begin{equation}
\xi=-\frac{l^{2}}{4\eta}=-\frac{3}{4\Lambda_{4}},
\end{equation} 
where $\eta=1$ for de Sitter and $-1$ for anti-de Sitter. This guarantees the vanishing of the boundary terms of the variation if the geometry is asymptotically (anti-)de Sitter\footnote{We thank Rodrigo Olea for pointing this out.}.

\section{Conclusions}\label{secv}

In this paper, we extended the validity of the entropy function formalism to the case of the extremal Kerr-(anti-)de Sitter black holes. The formalism generates the correct value of the Bekenstein-Hawking entropy. 
It generalizes the result in flat case. In $l \rightarrow \infty$ limit, we recover the result of Kerr black holes given in Ref.\cite{Astefanesei:2006dd} with $b=J/16 \pi$ (in $16\pi G_{N}=1$ unit). 
The result suggests the possibility of extending the rotating attractor to the case with the cosmological constant by including nontrivial scalar fields like the dilaton. Despite the non-supersymmetric background and the warped factorization of the near-horizon geometry, the entropy does not concern the asymptotic data of the geometry.

The way we showed the validity of the entropy formalism in the extremal Kerr-(anti-)de Sitter black holes was via Eq.(\ref{algerel}). The result is exactly the same as the one in Eq.(\ref{r1}) that was obtained from the black hole solution. Although the algebraic equation, being of fourth order in $F$, allows four roots in principle, only one of them corresponds to the value of $F$ in Eq.(\ref{entftn}). For example, one can see the relation gets truncated to second order in $F$ in the limit of $\lambda^{2}=\pi l^{2}\rightarrow\infty$, thus allows two different values $F=\pm 2\pi J$. Only the upper choice of the sign gives the value of $F$ in Eq.(\ref{entftn}), that is, $2\pi b$ in $l \rightarrow\infty$ limit.   

We also showed that the Gauss-Bonnet terms of the Lagrangian adds a constant correction to the entropy. Since the value is independent of any hair property of the black hole, we interpret its meaning just as the shift of the base point of the entropy. This shift is not physically sensible because what concerns us in some process is the change of the entropy rather than its value itself. 

This feature of the Gauss-Bonnet correction is peculiar in $4$-dimensions. Being topological in nature, the Gauss-Bonnet term of the Lagrangian does not affect the equation of motion either. 

The universal value of the shift in the entropy can be understood with the same reasoning. The value $\Delta F=2\pi\xi$ in Eq.(\ref{shift}) is independent of the angular momentum. This is because the topological quantity $\Delta F$ concerns the topology of the near horizon geometry, AdS$_{2}\times$S$^{2}$, which is the same for all extremal black holes in $4$-dimensions. (We also have the same topology in the stationary case where the factorization AdS$_{2}\times$S$^{1}$ is warped by a function of $\theta$.) The value is the same even for the black holes of Nariai class, where the near-horizon geometry is factorized as a two-dimensional de Sitter spacetime and a two-sphere \cite{Cho:2007we}. The topology of this latter near-horizon geometry is the same as that of the extremal black holes. This is clear in their isometries; $SO(1,2)$ for two-dimensional anti-de Sitter, and $SO(2,1)$ for two-dimensional de Sitter.

In this formalism of computing the black hole entropy, the extremality of the metric function is still important. The next mission would be to make this `extreme' condition looser. For example, we have to check the attractor behavior of the near-extremal black holes and finally of the Schwarzschild black hole. This issue will be treated elsewhere.

\appendix

\section{Signs and Ranges of Parameters}\label{apb}

\subsection{Determining the Choice of Sign in $\beta^2$}\label{apb1}
In Eq.(\ref{beta}), we have a multi-sign in its expression. The choice of the sign is determined by the conditions that $\beta^2>0$ and $p(\theta)=h^{2}(\theta)>0$. Eqs.(\ref{beta}) and (\ref{pth}) specify the conditions as 
\begin{eqnarray}
 \,6 b - {\eta\, l^2 } \pm \sqrt{l^4 + 48 b^2}&>& 0,\label{ineq1}\\
{{\beta^2 ( b\, l^2 + 3\eta \beta^2 )}} + {\eta\, b^{4}} \cos^2 \theta &>&0. \label{ineq2}
\end{eqnarray}

Let us analyze these conditions for de Sitter case and for anti-de Sitter case separately. 
\\
\\
\noindent (i) de Sitter case $(\eta=1)$ 

Since $l^4 + 48 b^2 > (6 b - l^2 )^2$, we must choose the upper sign of Eq.(\ref{ineq1}). This case satisfies Eq.(\ref{ineq2}) automatically so long as $b>0$. Therefore for de Sitter case, we have
\begin{eqnarray}\label{bs}
\beta^2 = {{ b} \over 6}\,\left(\, 6 b - l^2 + \sqrt{l^4 + 48 b^2}\,\right),\quad 0<b.
\end{eqnarray}

\noindent (ii) anti-de Sitter case $(\eta=-1)$ 

Eq.(\ref{ineq2}) gives a range for $\beta^2$;
\begin{eqnarray}\label{adsb}
 {b \over {6 }}\,\left(\,l^2 - \sqrt{l^4 - 12 b^2 \cos^2 \theta}\,\right)<\beta^2 <{b \over {6 }}\,\left(\,l^2 + \sqrt{l^4 - 12 b^2 \cos^2 \theta}\,\right).
\end{eqnarray}
To make sense of the square root, the parameter $b$ should be confined to the region $ 0< b< {{l^2}/{\sqrt{12}}}$. The range guarantees Eq.(\ref{ineq1}) for both signs. However, inserting Eq.(\ref{pth}) into Eq.(\ref{adsb}), we can refine the range further. We discard the upper sign of Eq.(\ref{beta}), otherwise Eq.(\ref{adsb}) cannot be satisfied.  As for the lower sign, Eq.(\ref{adsb}) can be satisfied as long as $b < {{l^2}/ 4}$. Summing up the result for the anti-de Sitter case, we have
\begin{eqnarray}
\beta^2 = {{ b} \over 6} \left(  6 b + {l^2 }  - \sqrt{l^4 + 48 b^2}\right),\quad 0< b<{{l^2}\over 4}.
\end{eqnarray}

\subsection{Ranges of Parameters $r_0^2$ and $b$}\label{apb2}
We have introduced two different parameters $r_0$ and $b$. The parameter $r_0$ appeared in Eq.(\ref{nhg2}) describing the horizon of an extremal Kerr-(anti)de Sitter black hole while $b$ appeared as an integration constant in the entropy function formalism. Comparing the near-horizon geometry of Eq.(\ref{a1}) (with the unknown functions determined in Eq.(\ref{rb}) via the entropy function formalism,) with the near-horizon geometry (\ref{nhg2}), one can find the following relation: 
\begin{eqnarray}\label{brds}
b = {{l^2  r_0^2\,( l^2 - 3 \eta r_0^2)}\over{l^4 - 6\, \eta\, l^2 r_0^2 - 3\,  r_0^4}}\,.
\end{eqnarray}
We showed in Eqs.(\ref{dsr})
 and (\ref{adsr}) that $r_0$ was restricted to be in a certain range: 
\begin{eqnarray}
\quad0\leq r_0^2 < \left({2 \over \sqrt{3}} -1 \right) l^2,\quad&&{\rm when} ~~\eta =1, \label{dsrb}\\ 
 \quad0 \leq r_0^2 <  {l^2 \over 3}, \quad &&{\rm when}~~\eta=-1 \label{adsrb}. 
\end{eqnarray}
We want to check whether these conditions on $r_0$ are consistent with the conditions on $b$ derived in appendix \ref{apb1}.
\\ 

\noindent (i) de Sitter case $(\eta = 1)$ 

From the range $b>0$, we have
\begin{eqnarray}
{{l^2  r_0^2\,( l^2 - 3 r_0^2)}\over{l^4 - 6\, l^2 r_0^2 - 3\,  r_0^4}} > 0.
\end{eqnarray}
Eq.(\ref{brds}) gives either 
\begin{eqnarray}
l^2 - 3 r_0^2 > 0, \quad &{\rm and}& \quad l^4  - 6 l^2 r_0^2 - 3 r_0^4 > 0, \label{rd1} \\
{\rm or},\quad l^2 - 3 r_0^2 < 0, \quad &{\rm and}& \quad l^4  - 6 l^2 r_0^2 - 3 r_0^4 < 0. \label{rd2}
\end{eqnarray}
Eq.(\ref{rd1}) gives $0 \leq r_0^2 < \left( {2 \over {\sqrt{3}}} -1 \right)l^2$, which is exactly the same as  Eq.(\ref{dsrb}). On the other hand, Eq.(\ref{rd2}) gives $r_0^2 > {l^2 / 3}$, which corresponds to unphysical solution as can be seen clearly in Fig. \ref{gds}(b).
\\

\noindent (ii) anti-de Sitter case $(\eta = -1)$ 

The allowed range of $b$ was  $0<b<l^2/4$. This leads to the range on $r_0^2$, $0 \leq r_0^2 < {l^2 \over 3}\,$, which is exactly the range we have as in Eq.(\ref{adsrb}). Therefore we conclude that the relation between $r_0$ and $b$ as given in (\ref{brds}) is correct.


\acknowledgments
This work is supported by the SRC program of KOSEF through CQUeST with grant number R11-2005-021 and the Korea Research Foundation Grant  funded by the Korean Government(MOEHRD) (KRF-2007-314-C00056).


\providecommand{\href}[2]{#2}\begingroup\raggedright\endgroup

\end{document}